\documentstyle{lamuphys}
\begin{document}
\renewcommand{\theequation}{\arabic{section}.\arabic{equation}}
\setcounter{equation}{0}


\def\wh{\widehat}
\def\wt{\widetilde}
\def\D{\Cal{D}}
\def\ov{\overline}
\def\un{\underline}
\def\noi{\noindent}


\title{On Some Stability Properties of Compactified
D=11 Supermembranes}

\author{I. Martín\inst{1} \and A.
Restuccia\inst{1} }

\institute{Universidad Simón Bolívar, Departamento de
Física\\Caracas 89000, Venezuela.\\e-mail:isbeliam@usb.ve,
arestu@usb.ve}

\maketitle

\begin{abstract}
We desribe the minimal configurations of the bosonic membrane
potential, when the membrane wraps up in an irreducible way over
$S^{1}\times S^{1}$. The membrane 2-dimensional spatial world
volume is taken as a Riemann Surface of genus $g$ with an
arbitrary metric over it. All the minimal solutions are obtained
and described in terms of 1-forms over an associated $U(1)$ fiber
bundle, extending previous results. It is shown that there are no
infinite dimensional valleys at the minima.
\end{abstract}

\section{Introduction}
The Minkowski $D=11$ Supermembrane, when the $SU(N)$, $N
\rightarrow \infty$, regularization is used, was shown to have a
continuous spectrum from zero to infinity (\cite{N + de Wit}). The
instability problem may be understood as a consequence of the
existence of string like configuration with the same energy. The
configurations which give the minimum of the potential consists of
infinite dimensional functional subspaces. The potential as a
functional may be described around those subspaces as having
infinite dimensional valleys. This property appears already in the
bosonic membrane, however because of the quantum zero point energy
of the oscillators transversal to the valleys stability is
attained. In the supermembrane, because of the property of the
$SUSY$ harmonic oscillators to have no zero point energy the
theory is unstable. Because of duality arguments and the relation
between the supermembrane and its dual D-brane, we are really more
interested in studying the Hamiltonian of the compactified
supermembranes where at least one dimension in the target space is
compactified to $S^{1}$ (\cite{Duff}). The spectrum of the
Hamiltonian of the compactified supermembrane has been recently
studied by several authors without a conclusive result
(\cite{Russo})(\cite{de Wit}).

In a recent paper (\cite{m+r+t})we showed that the Hamiltonian of
the membrane wrapped up in an irreducible way over $S^{1}\times
S^{1}$ has no infinite dimensional valleys. Moreover we found all
the minimal configurations, when the metric over $\Sigma$ the
2-dimensional spatial world volume of the membrane was the
canonical generalization of the induced metric over $S^{2}$ from
$R^{3}$. We considered the general situation where $\Sigma$ was a
Riemann Surface of genus $g$. The construction was performed in
terms of intrinsic harmonic coordinates. The minimal configuration
was found to be unique up to closed forms over $\Sigma$. We expect
that this properties of the bosonic Hamiltonian will also be valid
for the complete supermembrane Hamiltonian. That is, that there
are no infinite dimensional valleys at the minimal configurations
of the supermembrane wrapped up in an irreducible way over
$S^{1}\times S^{1}$. We will discuss this problem in a forthcoming
paper. In the present work we would like to give the general
solution for the minimal configurations when other metrics, not
necessarily the canonical one, is assumed over $\Sigma$.

\setcounter{equation}{0}
\section{Minimal Configurations of the
Membrane Potential}
\noindent

We will analyze in this section minimal configurations of the
Hamiltonian of the bosonic membrane. To each configuration of the
membrane, determined by $dx(\sigma)$ we will associate a
connection 1-form
 on a $U(1)$ bundle over $\sum$, a compact Riemann
surface of genus $g$. We will then show that the minimal
configurations correspond exactly to the magnetic monopoles over
Riemann surfaces found in (\cite{M+R})( \cite{F}). Given a set of
maps $X_{a}$, $a= 1,\ldots,9$, from $\sum$ to the target space $T$
we define
\begin{equation}
F_{ab}\equiv d(X_{a}dX_{b})  ,\, \, a,b = 1,\ldots,9.
\label{eq:2.1}
\end{equation}
as the $U(1)$ curvatures associated to the connection 1-forms
\begin{equation}
A_{ab}\equiv X_{[a}{dX_{b]}} \label{eq:2.2}
\end{equation}

The potential of the bosonic membrane may then be rewritten as
\begin{equation}
        <V>_{\sigma} \equiv \int_{\Sigma}d^{2}\sigma\sqrt{g}\,
V(\sigma)=
\int_{\Sigma}
{^\ast F_{ab} F_{ab}}
\label{eq:2.3}
\end{equation}
where ${^\ast F_{ab}}$ is the Hodge dual of the curvature 2-form
$F_{ab}$ and $g$ is the determinant of the metric over the Riemann
surface $\Sigma$. We normalize the metric by the condition $Vol \;
\Sigma=2 \pi$.

It is clear from (\ref{eq:2.3}) that the infinite dimensional
configuration
\begin{equation}
X_{a}(\sigma)={\lambda_{a}X(\sigma)}  ,\, \, a= 1,\ldots,9.
\label{eq:2.4}
\end{equation}

where $\lambda_{a}$ are arbitrary parameters, has zero potential.

In fact $F_{ab}=0$ for all $a$ and $b$. The configuration
(\ref{eq:2.4}) is infinite dimensional since $X(\sigma)$ is an
arbitrary map, with the only restriction to have a well defined
potential (\ref{eq:2.3}). The space of functions over $\Sigma$
satisfying that requirement is infinite dimensional. These are the
valleys which give rise to the continuous spectrum from zero to
infinite for the non-compactified supermembrane.

The existence of the valleys is, of course, not restricted to the
minima, as emphasized in (\cite{de Wit}).

We are interested in the analysis of the supermembrane in the case in
which the target space $T$ is compactified. Let $X(\sigma)$ denote a
compactified coordinate over $T$. Let us assume $X$ is a map from
\begin{equation}
\Sigma \rightarrow S^{1}
\label{eq:2.5}
\end{equation}

It is then straightforward to see that $dx$ satisfies the
following conditions for a 1-form $L$
\begin{eqnarray}
dL & = &0 \\ \nonumber
\oint_{{\cal C}_{i}}L & = & 2\pi n^{i}
\label{eq:2.6}
\end{eqnarray}
where $C_{i}$ denotes a basis of the integral homology of
dimension one over $\Sigma$. It is interesting to note that the
converse to (\ref{eq:2.6}) is also valid. That is, given a
globally defined 1-form $L$ over $\Sigma$ satisfying
(\ref{eq:2.6}) there exists a map $X$ from
\begin{equation}
\Sigma \rightarrow S^{1}
\label{eq:2.7}
\end{equation}
for which $L=dX$.

We will say the supermembrane wraps up in a non-trivial way when at
least one of the $n^{i}$ is different from zero.

We also say the supermembrane wraps up in an irreducible way over
$S^{1}\times S^{1}$ when
\begin{equation}
\int_{\Sigma}   dX \wedge dY \neq 0
\label{eq:2.8}
\end{equation}
where $X$ and $Y$ are two maps. That is, the irreducibility condition
requires a compactification of at least two coordinates on the
target $T$. If the membrane wraps up in an irreducible way over
$S^{1}\times
S^{1}$ it does it in a nontrivial way over each of the $S^{1}$.
Moreover
\begin{equation}
\frac{1}{2\pi}\int_{\Sigma}     dX \wedge dY =2\pi N \neq 0,
\label{eq:2.9}
\end{equation}
$N$ being an integer.

(\ref{eq:2.8}) may be interpreted in terms of the $U(1)$ bundle
associated to

the membrane configuration, using (\ref{eq:2.1}) and (\ref{eq:2.2}). It
tell
us that the
corresponding Chern class is non-trivial. The integral number $N$ in
(\ref{eq:2.9}) determines the $U(1)$ bundle over which the connection
(\ref{eq:2.2}) is
defined.  We notice that the existence of infinite dimensional
valleys still persist when the target space $T$ is of the form
\begin{equation}
T=M_{10}\times S^{1}.
\label{eq:2.10}
\end{equation}

In fact the coordinates mapping $\Sigma \rightarrow M_{10}$,
10-dimensional Minkowski space, are single valued over $\Sigma$.

We may always take the compactified coordinate, say $X_{1}$ to be
\begin{equation}
X_{1}=\phi
\label{eq:2.11}
\end{equation}
where $\phi$ is the angular coordinate of $S^{1}$ in (\ref{eq:2.10}).
An admissible configuration is then given by
\[
X_{2}=X_{3}=\ldots=X_{9}=\frac{dX(\phi)}{d\phi}
\]
\begin{equation}
\phi=\phi(\sigma),
 \label{eq:2.12}
\end{equation}
 where ${X(\phi)}$ is a differentiable single valued function on
 $\phi$, that is a regular periodic function, and $\phi(\sigma)$
 is a map from $\Sigma$ to $S^{1}$. It then follows that the
 curvature (\ref{eq:2.1}) is zero for all $a$ and $b$. The subspace
 (\ref{eq:2.12}) is still
infinite dimensional.

 We look now for the stationary points of (\ref{eq:2.3}) over the space
of
maps
defining supermembranes with irreducible wrapping over
$S^{1} \times S^{1}$. It is straightforward to see in this case  that
the
minimal configurations occur when all but $X,Y$ maps are zero.
Associated
to this space we may introduce an
$U(1)$ principle bundle.  We proceed by noting that

\begin{equation}
F=\frac{1}{2\pi}dX \wedge dY
\label{eq:2.13}
\end{equation}
is a closed 2-form globally defined over $\Sigma$ satisfying
(\ref{eq:2.9}). By Weil's Theorem (\cite{H}), (\cite{C+M+R}),
there exists a $U(1)$ principle bundle and a connection over it
such that its pull back by sections over $\Sigma$ are 1-form
connections with curvatures given by (\ref{eq:2.13}).

The stationary points of the potential satisfy
\begin{eqnarray}
\delta X dY \wedge d {^\ast F}& = &0 \\\nonumber
\delta Y dX \wedge d {^\ast F} &= &0
\label{eq:2.14}
\end{eqnarray}
which imply

\begin{equation}
d {^\ast F} = 0.
\label{eq:2.15}
\end{equation}

Now, since $*F$ is a 0-form we get

\[
^\ast F = constant.
\]
over $\Sigma$, and using (\ref{eq:2.9}) we finally obtain
\begin{equation}
^\ast F =\frac{2\pi N}{Vol\Sigma}=N.
\label{eq:2.16}
\end{equation}

We will now show that the configurations with $X$ and $Y$
satisfying (\ref{eq:2.16}) and all other $X_{a}$ maps to zero are
minima of the potential (\ref{eq:2.3}) within the space of
configurations with irreducible winding (\ref{eq:2.9}).

For any connection on the $U(1)$ principle bundle over $\Sigma$
determined by $N$ the associated curvature 2-form satisfies
\begin{eqnarray}
dF &= &0 \\ \nonumber
\int_{\Sigma} F&=&2\pi N \label{eq:2.17}
\end{eqnarray}

Let $A_{o}$ be a connection 1-form satisfying (\ref{eq:2.16}), and
$A_{1}$ any other connection 1-form on the sample principle
bundle. Then, using (\ref{eq:2.17}) for $F(A_{1})$, we obtain
\begin{equation}
\int_{\Sigma} {^\ast F(A_{1}-A_{0}) F(A_{1}-A_{0})}=
\int_{\Sigma}[^\ast F(A_{1}) F(A_{1}) - ^\ast F(A_{0}) F(A_{0})].
\label{eq:2.18}
\end{equation}

The left hand member of (\ref{eq:2.18}) is greater or equal to
zero, we then have
\begin{equation}
\int_{\Sigma} {^\ast} F(A_{1}) F(A_{1}) \geq \int_{\Sigma} {^\ast}
F(A_{0}) F(A_{0}) \label{eq:2.19}
\end{equation}
The equality in (\ref{eq:2.19}) is obtained when the left hand member of

(\ref{eq:2.18})
is zero. This implies

\begin{equation}
{^\ast F(A_{1}-A_{0})}= ^\ast F(A_{1}) - ^\ast F(A_{0})=0.
\label{eq:2.20}
\end{equation}

That is, the equality in (\ref{eq:2.19}) is obtained if and only if
\begin{equation}
A_{1}=A_{0}+d\Lambda
\label{eq:2.21}
\end{equation}
where $d\Lambda$ is a closed 1-form globally defined over $\Sigma$.

The space of regular closed 1-forms, modulo exact 1-forms, over a
compact (closed) Riemann Surface is finite dimensional. The exact
1-forms correspond to gauge transformations on the $U(1)$ bundle.
In (\cite{m+r+t})it was shown that they are generated by the area
preserving transformations on the membrane maps. This implies the
non-existence of infinite dimensional valleys at the minima for
the membranes wrapping up in an irreducible way onto $S^{1}\times
S^{1}$.

\setcounter{equation}{0}
\section{Minimal Connections: Magnetic
Monopoles over Riemann Surfaces of genus $g$}
\noindent

We will show in this section how to construct all the minimal
connections over $S^{2}$ and over all topologically non-trivial
Riemann Surfaces. To do so we will construct one minimal connection
for each $N$. All others are obtained from (\ref{eq:2.21}). The space of

closed
1-forms modulo exact forms is the space of harmonic 1-forms over
$\Sigma$. It has been extensively studied in the literature, so it is
not necessary to discuss it there. Our problem reduces then to find
one minimal connection for each $N$, That is for each principle bundle
over $\Sigma$. We will describe now that construction.

The explicit expression of the monopole connections is obtained in terms
of
the
abelian differential $d\tilde{\Phi}$ of the third kind over the
compact Riemann surface $\Sigma$ of genus g.  $d\tilde{\Phi}$ is a
meromorphic 1-form with poles of residue +1 and -1 at points $a$ and
$b$,
with real normalization. $\tilde{\Phi}$ is the abelian integral,
its real part $G(z,\bar{z},a,b,t)$ is a harmonic univalent function
over $\Sigma$ with logarithmic behavior around $a$ and $b$

\begin{eqnarray}
        \ln(\frac{1}{|z+ a|}) & +& \mbox{regular terms ,}  \nonumber \\
        \ln{|z-b|} & + & \mbox{regular terms ,}
        \label{eq:3.1}
\end{eqnarray}

It is a conformally invariant geometrical object. $z$ denotes the
local coordinate over $\Sigma$ and $t$ the set of $3g-3$
parameters describing the moduli space of Riemann surfaces. We are
considering maps from $\Sigma \mapsto S^{1}\times S^{1}\times
M^{7}$ for a given $\Sigma$, so the parameters $t$ are kept fixed.
They show however that the construction of minimal connections is
a conformally invariant one.

Let $a_{i}$, $i=1,...,m$ be m points over the compact Riemann
surface.  We associate to them integer weights $\alpha_{i}$,
$i=1,...,m$, satisfying

\begin{equation}
\sum_{i=1}^{m} \alpha _{i} =0
\label{eq:3.2}
\end{equation}

We define
\begin{equation}
\phi = \sum_{i=1}^{m} \alpha _{i} G(z,\bar{z},a_{i},b,t).
\label{eq:3.3}
\end{equation}
and  have
\begin{eqnarray*}
        \phi  & \rightarrow & - \infty \,\, \mbox{at $a_{i}$ with negative weights}\\
        \phi     & \rightarrow & + \infty \,\, \mbox{ at $a_{i}$ with positive weights}.
\end{eqnarray*}
$\alpha_{i}$ are integers in order to have univalent transition
functions over the nontrivial fiber bundle that we consider.

We denote $\tilde{\Phi}$ the abelian integral with real part
$\phi$.  Its imaginary part $\varphi$ is also harmonic but
multivalued over $\Sigma$,
\begin{equation}
\tilde{\Phi}    = \phi +i\varphi.
\label{eq:3.4}
\end{equation}

Let us consider the curve ${\cal {C}}$ over $\Sigma$ defined by

\[
\phi =constant.
\]

It is a closed curve homologous to zero.  It divides the Riemann
surface into two regions $U_{+}$ and $U_{-}$, where $U_{+}$ contains
all the points $a_{i}$ with negative weights and $U_{-}$ the ones
with positive weights.

We define over $U_{+}$ and $U_{-}$ the connection 1-forms
\begin{eqnarray}
A_{+} & = & \frac{1}{2}(1+\tanh(\phi)) d\varphi \nonumber \\
A_{-}&= &\frac{1}{2}(-1+ \tanh(\phi))d\varphi \label{eq:3.6}
\end{eqnarray}
respectively.
$A_{+}$ is regular in $U_{+}$ and $A_{-}$ in $U_{-}$.
In the overlapping $U_{+} \bigcap U_{-}$ we have
\begin{equation}
A_{+}=A_{-}+d\varphi
\label{eq:3.7}
\end{equation}
$g= exp ({i\varphi})$ defines the transition function on the overlapping

$U_{+} \bigcap U_{-}$, and because of the integer weights it is
univalued over $U_{+} \bigcap U_{-}$.

The base manifold $\Sigma$, the transition function $g$ and the
structure group $U(1)$ have a unique class of equivalent $U(1)$
principle bundles over $\Sigma$ associated to them. (3.6) defines
a 1-form connection over $\Sigma$ with curvature

\begin{equation}
F=\frac{1}{2}\frac{1}{\cosh{^2}{\phi}}d\phi\wedge d\varphi
\label{eq:3.8}
\end{equation}

The $U(1)$ principle bundles are classified by the sum of the
positive integer weights $\alpha _{i}$

\begin{equation}
N= \sum_{i}\alpha_{i}^{+}  \; \; , \alpha_{i}^{+} > 0.
\label{eq:3.9}
\end{equation}
which is the only integer determining the number of times
$\varphi$ wraps around ${\cal {C}}$. All the bundles with the same
N are equivalent. (\ref{eq:3.8}) satisfies (2.17), moreover it
also satisfies (\ref{eq:2.16}). In fact, since $\varphi$ and
$\phi$ are harmonic over $\Sigma$, the metric is

\begin{equation}
d^{2}s=\frac{1}{\cosh ^{2} {\phi} }((d\varphi)^{2} + (d\phi)^{2}),
\label{eq:3.10}
\end{equation}
and then (\ref{eq:2.16}) follows directly.

We have then found for each $U(1)$ principle bundle over $\Sigma$,
a connection 1-form (3.6) with curvature 2-form (\ref{eq:3.8})
satisfying (\ref{eq:2.16})for the metric (\ref{eq:3.10}) on the
Riemann Surface. We have obtained several expressions
(3.6),(\ref{eq:3.8}) and (\ref{eq:3.10}), since we are allowed to
consider different harmonic coordinates $\phi$ and $\varphi$. In
fact, for any set of $a_i$ with total positive weight $N$, we may
define coordinates $\phi$ and $\varphi$ away from the points $a_1$
and $b$. By so doing we are only using different coordinates over
$\Sigma$ to describe the same connection over the $U(1)$ principle
bundle. It is, as if in the expressions of the 1-form connection
describing the Dirac monopole we use different coordinates
$(\theta,\varphi)$ with different North and South poles to
describe the magnetic field of the monopole. Since $^\ast F$ is
scalar field over $\Sigma$ then it is always equal to $N$.
\setcounter{equation}{0}
\section{The General Solution}
 \noindent
We have thus obtained all the solutions satisfying $^\ast F=N$ for
the metric (\ref{eq:3.10}). The question arises then, what happens
when we consider, instead of (\ref{eq:3.10}), the metric
\begin{equation}
|\lambda(\phi
,\varphi)|^{2}\frac{1}{\cosh^{2}(\phi)}[(d\varphi)^{2}+(d\phi)^{2}]
\label{eq:4.1}
\end{equation}
that is, an element of the conformal class of (\ref{eq:3.10}). The
questions is relevant since (\ref{eq:2.16}) is not conformal
invariant. It depends on the metric through the factor
$(\sqrt{g})^{-1}$. We may then ask for the solutions of
(\ref{eq:2.16}) when the new metric (\ref{eq:4.1}) is considered
over $\Sigma$. We will answer the question starting with the case
in which $\Sigma$ is the sphere $S^2$, and giving afterwards the
solutions for topologically non-trivial Riemann Surfaces.

We consider the Hopf fiber bundle over $S_2$. The three
dimensional sphere $S_3$ may be defined by $z_{0},z_{1}\in C$, the
complex numbers, satisfying
\begin{equation}
z_{0}\bar{z}_{0}+z_{1}\bar{z}_{1}=1. \label{eq:4.2}
\end{equation}
The group $U(1)$ acts on $S_{3}$ by
\begin{equation}
\left( z_{0},z_{1}\right) \rightarrow \left( z_{0}u ,z_{1}u
\right) \label{eq:4.3}
\end{equation}
where $u\bar{u}=1$, $\bar{u}$ being the complex conjugate to $u\in
C$. The projection $S_{3}\rightarrow S_{2}$ is defined by the
composition of
\begin{equation}
\left( z_{0},z_{1}\right) \rightarrow \left\{
\begin{array}{c}
\frac{z_{1}}{z_{0}}\;\;\;z_{0}\neq 0 \\
\frac{z_{0}}{z_{1}}\;\;\;z_{1}\neq 0
\end{array}
\right. \label{eq:4.4}
\end{equation}
with the stereographic projection
\[
C \rightarrow S_{2}
\]
defined by
\begin{equation}
\rho=\frac{\sin(\theta)}{1-\cos(\theta)} \label{eq:4.5}
\end{equation}
where $z=\rho e^{i\phi}\in \; C$ and $(\theta ,\phi)$ are the
coordinates $S_2$.

There is a natural connection over the Hopf fiber bundle which may
be obtained from the line element of $S_3$,
\begin{equation}
ds^{2}=4\left(d\bar{z}_{0}dz_{0}+d\bar{z}_{1}dz_1 \right)
\label{eq:4.6}
\end{equation}
where $z_{0},z_{1}$ satisfy (\ref{eq:4.2}). We will use spherical
coordinates $(\chi,\theta,\phi)$ over $S_3$, defined in the
following way
\begin{eqnarray}
z_{0}&=&\exp\left[
\frac{1}{2}i(\chi+\varphi)\right]C(\theta,\varphi)\\ \nonumber
z_{1}&=&\exp\left[
\frac{1}{2}i(\chi-\varphi)\right]S(\theta,\varphi) \label{eq:4.7}
\end{eqnarray}
where
\begin{equation}
C^{2}+S^{2}=1.
\end{equation}

We then obtain the line element of $S_3$ as
\[
\frac{1}{4}ds^{2}=(dc)^{2}+(ds)^{2}+\frac{1}{4}\left[(d\chi)^{2}+(d\varphi)^{2}+
2(c^{2}-s^{2}) d\chi d\phi \right].
\]
We denote
\begin{eqnarray*}
c^{2}-s^{2}&\equiv& g(u,\varphi)\\
 u&\equiv&\cos(\theta)
\end{eqnarray*}
We then get
\begin{equation}
ds^{2}=(d\chi + g(u,\varphi)d\phi)^{2} + \frac{(dg)^{2}}{1-g^{2}}+
(1-g^{2})(d\varphi)^{2}. \label{eq:4.9}
\end{equation}
We notice that in the particular case
\begin{eqnarray}
C(\theta , \varphi)&=&\cos(\frac{\theta}{2})\\ \nonumber S(\theta
, \varphi&=& \sin(\frac{\theta}{2})) \label{eq:4.10}
\end{eqnarray}
(\ref{eq:4.9}) yields
\begin{equation}
ds^{2}=(d\chi+ \cos(\theta)d\varphi)^{2}+ (d\theta)^{2} +
(\sin(\theta)d\varphi)^{2}. \label{eq:4.11}
\end{equation}

Coming back to the general case (\ref{eq:4.9}), the line element
of $S_3$ decomposes into the line element of $S_2$
\begin{equation}
\frac{(dg)^{2}}{1-g^{2}}+(1-g^{2})(d\varphi)^{2} \label{eq:4.12}
\end{equation}
and the tensorial square of the 1-form
\begin{equation}
\omega=d\chi+g(u ,\varphi)d\varphi. \label{eq:4.13}
\end{equation}
The above decomposition allows to determine a 1-form
(\ref{eq:4.13}) over the fiber bundle $S_3$. Notice that from
(\ref{eq:4.3}) and (4.7) the group acts on $\chi$ as follows
\begin{equation}
\chi \rightarrow \chi+\lambda \label{eq:4.14}
\end{equation}
where $u=\exp(i\lambda)$. We are then interested in considering
\[
\frac{1}{2}\omega
\]
as a connection over the fiber bundle $S_3$.

To obtain the $U(1)$ connection 1-form over $S_2$, one may
consider the local section
\begin{eqnarray}
\tilde{z}_{0}&=&\exp(i\varphi)C(\theta , \varphi)\\ \nonumber
\tilde{z}_{1}&=&S(\theta , \varphi) \label{eq:4.15}
\end{eqnarray}
over $S_2$ with the point $\theta=0$ removed, which we denote
$U_+$. The $U(1)$ connection over $U_+$ is then
\begin{equation}
A_{+}=\frac{1}{2}(1+g(u,\varphi))d\varphi. \label{eq:4.16}
\end{equation}
To give a covering of $S_2$ we define $U_{-}$, another local
section, by
\begin{eqnarray}
\hat{z}_{0}&=&C(\theta , \varphi) \\\nonumber \hat{z}_{1}&=&
e^{-i\varphi} S(\theta , \varphi) \nonumber \label{eq:4.16a}
\end{eqnarray}
over $S_2$ with the point $\theta=\pi$ removed. We have assumed in
(4.15) and (4.17) that
\[
\left. C \right|_{\theta=\pi}=0
\]
\begin{equation}
\left.S\right|_{\theta=0}=0. \label{eq:4.17}
\end{equation}
Over $U_{-}$ the connection 1-form is then given by
\begin{equation}
A_{-}=\frac{1}{2}(g(u , \varphi) -1)d\varphi. \label{eq:4.18}
\end{equation}
In the overlapping region $U_{+}\cap U_{-}$ we have
\begin{equation}
A_{+}-A_{-}=d\varphi. \label{eq:4.19}
\end{equation}
The curvature 2-form $F$ is then given by
\begin{equation}
F=\frac{1}{2}\partial_{\mu}g(u, \varphi)\sin(\theta)d\varphi
\wedge d\theta. \label{eq:4.20}
\end{equation}
In the particular case (\ref{eq:4.10}), it reduces to
\[
F=\frac{1}{2}\sin(\theta)d\varphi \wedge d\theta.
\]

In order to obtain the general solution satisfying (2.16) we may
proceed in two ways. We may extend the Hopf fibring
\[
S_{3} \rightarrow S_{2}
\]
to
\[
S_{2n+1} \rightarrow CP_{n}
\]
as considered in (\cite{Trautman}), and repeat the procedure. This
approach yields the explicit expressions of the connection 1-form
over $CP_{n}$ and then over $S_{2}$. Otherwise we may consider
\begin{equation}
F=N\frac{1}{2}\partial_{mu}g(u , \varphi) \sin(\theta) d\varphi
\wedge d\theta \label{eq:4.21}
\end{equation}
and check (2.17). Weil's theorem ensures the existence of the
connection over a $U(1)$ fiber bundle over $\Sigma$, with a
curvature 2-form given by (\ref{eq:4.21}). This second procedure,
although more direct, does not provide the explicit expression of
the connection 1-form as in (\ref{eq:4.16}) and (4.19).

If is straightforward to check that (\ref{eq:4.21}) satisfies
(2.17). In fact
\begin{eqnarray}
C^{2}&=& \frac{1+g}{2}\\\nonumber
S^{2}&=&\frac{1-g}{2}\nonumber
\end{eqnarray}
and hence at $\theta=\pi$, g=-1 and at $\theta=0$, $g=1$. We may
now evaluate $^\ast F$ for our general solution (\ref{eq:4.21})
and the metric (\ref{eq:4.12}).

If we use our normalization $Vol\Sigma=2\pi$, we obtain
\[
\sqrt{g}=\frac{1}{2}\partial_{\mu}g\sin(\theta),
\]
and
\begin{equation}
^\ast F=\frac{2}{\sqrt{g}}F_{\varphi\theta}=N
\label{eq:4.22}
\end{equation}
as required for the minima of the membrane potential. The above
construction may be extended to obtain all the minimal connection
1-form over topologically non-trivial Riemann surfaces. Using the
global coordinates introduced in section 3, the connection 1-form
over $U_{+}$ and $U_{-}$ may be expressed as
\begin{eqnarray}
A_{+}&=&\frac{N}{2}\left( 1+ g(u,
\varphi)\right)d\varphi,\\\nonumber
 A_{-}&=&\frac{N}{2}\left( -1+
g(u, \varphi)\right)d\varphi, \label{eq:4.23}
\end{eqnarray}
respectively, where $u=\tanh(\phi)$ and $g(u , \varphi)$ is a
single valued function over $\Sigma$ satisfying
\begin{eqnarray}
u \rightarrow +1 && g \rightarrow +1 \\ \nonumber
u \rightarrow -1&& g \rightarrow -1. \nonumber
\end{eqnarray}
The curvature 2-form has then the form
\begin{equation}
F=N\frac{1}{2}\partial_{\mu}g(u ,
\varphi)\frac{1}{\cosh^{2}(\theta)} d\phi \wedge d\varphi .
\label{eq:4.24}
\end{equation}
Its Hodge dual, over the metric (\ref{eq:4.12}), is consequently
\begin{equation}
^\ast F=N.
\label{eq:4.25}
\end{equation}
(4.25) gives then the general solution for the minimization
problem of the membrane potential over topologically non-trivial
Riemann Surface.

Having constructed all minimal configurations of the membrane
potential, in terms of connection 1-forms over $U(1)$ fiber
bundles, one has to determine the configurations maps in terms of
them. If $\hat{A}$ is a minimal connection, then under an area
preserving diffeomorphism
\[
\delta \hat{A}_{r}=
\partial_{r}(-\epsilon^{st}\partial_{t}\xi\hat{A}_{s}-
\frac{1}{2}\xi{^\ast}\hat{F})
\]
where $\xi$ is the infinitesimal parameter of the transformation.
It is then equivalent to a gauge transformation on the $U(1)$
fiber bundle. Using this property it was shown in (\cite{m+r+t})
that the space of all minimal connections may be generated from a
particular minimal connection to which a representative of each
real cohomology class of 1-form over $\Sigma$ has been added. The
space of maps given rise to the particular minimal connection
being finite dimensional. The argument in (\cite{m+r+t}) was
performed using the canonical metric over $\Sigma$, its extension
to the general case is straightforward.

\end{document}